\documentclass[10pt]{article} 
\pdfoutput=1

\usepackage{cite} 
\usepackage{url}  
\usepackage{ifthen}  
\usepackage{multicol}   
\usepackage[utf8]{inputenc} 
\renewcommand{\thefootnote}{\fnsymbol{footnote}}
\usepackage{rotating}
\usepackage{amstext}
\usepackage{epsfig}
\usepackage{amsmath, amsthm, amssymb}
\usepackage{graphicx}
\usepackage{cite}
\def\be{\begin{equation}}
\def\ee{\end{equation}}
\def\bq{\begin{eqnarray}}
\def\eq{\end{eqnarray}}
\def\beq{\begin{eqnarray}}
\def\eeq{\end{eqnarray}}

\newcommand{\prob}{\mathbf{P}}
\newboolean{publ}

\begin{document}


\title{Continuous time Boolean modeling for biological signaling: application of Gillespie algorithm}
 

\author{Gautier Stoll
         (gautier.stoll@curie.fr)\\ Eric Viara (viara@sysra.com)\\
         Emmanuel Barillot (emmanuel.barillot)\\ Laurence Calzone (laurence.calzone@curie.fr%
      }



\maketitle


\begin{abstract}
This article presents an algorithm that allows modeling of biological networks in a qualitative framework with continuous time. Mathematical modeling is used as a systems biology tool to answer biological questions, and more precisely, to validate a network that describes biological observations and to predict the effect of perturbations.

There exist two major types of mathematical modeling approaches: (1) Quantitative modeling, describing various chemical species concentrations by real numbers, mainly based on differential equations and chemical kinetics formalism; (2) Qualitative modeling, describing chemical species concentrations or activities by a finite set of values, the most used being the Boolean modeling. 

One major critics of qualitative modeling is that time in this approach is represented by discrete steps. Qualitative approaches, on the other hand, can efficiently describe and predict stable states, but remain inconvenient in describing transient kinetics leading to these states. In order to handle these transient events, several attempts were made to introduce continuous time in qualitative modeling. They consist in refining the time evolution in this formalism by introducing priority classes (some variables are updated with a higher priority), variable time delays, etc.

Here, we propose a modeling approach that is intrinsically continuous in time. The algorithm presented here fills the gap between qualitative and quantitative modeling. It is based on continuous time Markov process applied on a Boolean state space. In order to describe the temporal evolution, we explicitly specify the transition rates for each node. For that purpose, we built a language that can be seen as a generalization of Boolean equations. The values of transition rates have a natural interpretation: it is the inverse of the time for the transition to occur. Mathematically, this approach can be translated in a set of ordinary differential equations on probability distributions; therefore, it can be seen as an approach in between quantitative and qualitative.

We developed a C++ software, MaBoSS, that is able to simulate such a system by applying Kinetic Monte-Carlo (or Gillespie algorithm) in the Boolean state space. This software, parallelized and optimized, computes temporal evolution of probability distributions and can also estimate stationary distributions.  
Applications of Boolean Kinetic Monte-Carlo have been demonstrated for two qualitative models: a toy model and a published p53/Mdm2 model. Our approach allows to describe kinetic phenomena which were difficult to handle in the original models. In particular, transient effects are represented by time dependent probability distributions, interpretable in terms of cell populations.
       
\end{abstract}

\ifthenelse{\boolean{publ}}{\begin{multicols}{2}}{}



\section*{Background}

Mathematical models of signaling pathways can be seen as tools to answer biological questions. The most widely used mathematical formalisms to answer these questions are ordinary differential equations (ODEs) and Boolean modeling.

Ordinary differential equations (ODEs) have been widely used to model signaling pathways. It is the most natural formalism for translating detailed reaction networks into a mathematical model. Indeed, equations can be directly derived using mass action laws, Michaelis-Menten kinetics or Hill functions for each reaction in order to account for the observed behaviors. This framework has limitations, though. The first one concerns the difficulty of assigning the kinetic parameter values in the model. Ideally, these parameters would be extracted from experimental data. However, they are often chosen so as to fit qualitatively the expected phenotypes.
The second limitation arises when studying cell population heterogeneity. In this case, ODEs are no longer appropriate since the approach is deterministic and thus focuses on the average behavior. To include non-determinism, an ODE model needs to be transformed into stochastic chemical model. In this formalism, a master equation is written on the probabilities of number of molecules for each species (instead of being written on ODEs of continuous concentrations). In the translation process, the same parameters used in ODEs (more particularly in ODEs written with mass action law) can be used in the master equation, but in this case, the number of initial conditions explodes along with the computation time.

Boolean formalism is another formalism used to model signaling pathways where genes/proteins are parameterized by 0s and 1s only. It is the most natural formalism to translate an influence network into a mathematical model. In such networks, each node corresponds to a species and each arrow to an interaction or an influence (positive or negative). In a Boolean model, a logical rule integrating the signs of the input links is assigned to each node. As a result, there are no real parameter values to adjust besides choosing the appropriate logical rules that best describe the system. In this paper, we will refer to a network state as a state in which each node of the influence network has a Boolean value. The set of all possible transitions between the network states is defined as a transition graph. There are two types of transition graphs, one deduced from the synchronous update strategy \cite{kauffman1969homeostasis}, for which all the nodes that can be updated are updated in one transition, and another one deduced from the asynchronous update strategy \cite{thomas1991regulatory}, for which only one node is updated in one transition. In the Boolean formalism, each transition can be interpreted as a time step. Real characterization of biological time is not taken into consideration. However, in many biological problems, time plays a crucial role. 

As mentioned for ODE, stochasticity is an important aspect when studying cell populations, It can be done: on nodes (by randomly flipping the node state \cite{stoll2006few,stoll2010stabilizing}), on the logical rules (by allowing to change an AND gate by an OR gate \cite{garg2009modeling}), and on the update rules (by defining the probability and the priority of changing one Boolean value in an asynchronous strategy \cite{faure2006dynamical} or by adding noise to the whole system in a synchronous strategy \cite{shmulevich2002probabilistic}). One of the main drawbacks of the Boolean approach is the explosion of solutions. In an asynchronous update strategy, the transition graph can reach $2^\text{\#nodes}$.  

Both discrete and continuous frameworks have advantages and disadvantages above-mentioned. We propose here to combine some of the advantages of both approaches with what we call the ``Boolean Kinetic Monte-Carlo'' algorithm (BKMC). It consists of a natural generalization of the asynchronous Boolean dynamics \cite{thomas1991regulatory}, with a direct probabilistic interpretation. In BKMC framework, the dynamics is parameterized by a biological time, the order of update is noisy (less strict than priority classes introduced in GINsim \cite{gonzalez2006ginsim}). A BKMC model is specified by logical rules as in regular Boolean models but with a more precise information: a numerical rate is added for each transition. 

BKMC is best suited to model signaling pathways in the following cases:
\begin{itemize}
\item The model is based on an influence network, because BKMC is a generalization of the asynchronous Boolean dynamics. Typical examples are published models of cell cycle \cite{faure2006dynamical}. 
\item The model describes processes for which information about biological time is known (relative rate, order, etc.), because in BKMC, time is parameterized by a real number. This is typically the case when studying developmental biology, where animal models provide time changes of gene/protein activities \cite{wunderlich2011modeling}. 
\item The model describes heterogeneous cell population behavior, because BKMC has a probabilistic interpretation. For example, modeling heterogeneous cell population can help to understand tissu formation based on cell differentiation \cite{macarthur2009systems}. 
\item The model can contain many nodes (up to 64 in the present implementation), because BKMC is a simulation algorithm that converges fast. This can be useful for big models that have already been modeled with a discrete time Boolean method \cite{saez2011comparing}, in order to obtain a finer description of transient effects.
\end{itemize}

Previous published works have also introduced a continuous time approach in the Boolean framework(\cite{siebert2008temporal}, \cite{oktem2002computational}, \cite{vahedi2009sampling}, \cite{sevim2010reliability}, \cite{teraguchi2011stochastic}, \cite{bauer2010receptor}, \cite{p53Model}).
We will review them and present BKMC approach. We will describe the C++ software, MaBoSS, developed to implement BKMC algorithm and illustrate its use with two examples, a toy model and a published model of p53-MDM2 interaction.

All abbreviations, definitions and algorithms used in this article can be found in Supplementary Material. Throughout the article, all terms that are italicized are defined in the Supplementary Material (Part 3).

\section*{Results and discussion}
\subsection*{BKMC for continuous time Boolean model}

\subsubsection*{Continuous time in Boolean modeling: past and present} 

In our context, we briefly recall that, in Boolean approaches for modeling networks, we define the state of each node of the network by a Boolean value (node state) and the network state by the set of node states. Any dynamics in the transition graph is represented by sequences of network states. A node state is based on the sign of the input arrows and the logic that links them. The dynamics can be deterministic (in the case of synchronized update \cite{kauffman1969homeostasis}) or non-deterministic (in the case of asynchronized update \cite{thomas1991regulatory} or probabilistic Boolean networks \cite{shmulevich2002probabilistic}).

The difficulty to interpret a dynamics in terms of biological time has led to several works that have generalized Boolean approaches. These approaches can be divided in two classes that we call explicit and implicit time for discrete steps.

The explicit time for discrete steps consists of adding a real parameter to each node. These parameters correspond to the time associated to each node state before it flips to another one. They need to be set for each node state (\cite{oktem2002computational}, \cite{siebert2008temporal}). Because data about these time lengths are difficult to extract from experimental studies, some works have included noise in the definition of these time parameters \cite{p53Model}. The drawback of this method is that the computation of the Boolean model becomes sensitive to both the type of noise and the initial conditions. As a result, these time parameters become new parameters that need to be tuned carefully and thus add complexity to the modeling. 

The implicit time for discrete steps consists of adding probabilities to each transition of the transition graph, in the case of non-deterministic transitions. It is argued that these probabilities could be interpreted as adding time to a biological process. As an illustration, let us assume a small network of two nodes, A and B. At time t, A and B are inactive: [AB] = [00]. In the transition graph, there exist two possible transitions at t+1: [00] $\rightarrow$ [01] and [00] $\rightarrow$ [10]. If the first transition has a significant higher probability than the second one, then we can conclude that B has a higher tendency to activate before A. Therefore, it is equivalent to say that the activation of B is faster than the activation of A. Thus, in this case, the notion of time is implicitly modeled by setting probability transitions. In particular, \emph{priority rules}, in the asynchronous strategy, consist of putting some of these probabilities to zero \cite{faure2006dynamical}. In our example, if B is faster than A then the probability of the transition [00] $\rightarrow$ [10] is zero. As a result, the prioritized nodes always activate before the others. From a different perspective but keeping the same idea, Vahedi and colleagues \cite{vahedi2009sampling} have set up a method to deduce explicitly these probabilities from the duration of each discrete step. 
With the implementation of implicit time in a Boolean model, the dynamics remains difficult to interpret in terms of biological time.

As an alternative to these approaches, we propose BKMC algorithm.


\subsubsection*{Properties of BKMC algorithm}

BKMC algorithm was built such as to meet the following principles:
\begin{itemize}
\item The state of each node is given by a Boolean number (0 or 1) (referred to as node state); 
\item The state of the network is given by the set of node states (referred to as network state); 
\item The update of a node state is based on the signs linking the incoming arrows of this node and the logic;
\item Time is represented by a real number;
\item Evolution is stochastic.
\end{itemize}

For that, we choose to describe the time evolution of network states by a {\bf Markov process} with {\bf continuous time}. Therefore, the dynamics is defined by {\bf transition rates} inserted in a master equation (see Supplementary Material, section 1.1). Transitions for which a rate is specified are based on asynchronous Boolean dynamics. 

\subsubsection*{Markov process for Boolean model}

In BKMC, we adapt the Markov process to the Boolean approach.

Consider a network of $n$ nodes (or agents, that can represent any species, \emph{i.e.} mRNA, proteins, complexes, etc.). In a Boolean framework, the network state of the system is described by a vector ${\bf S}$ of Boolean values, i.e. $S_i \in \{0,1\}, i=1,\ldots,n$ where $S_i$ is the {\sl state of the node} $i$. The set of all possible network states, also referred to as the {\sl network state space}, will be called $\Sigma$. 

A stochastic description of state evolution is represented by a {\sl stochastic process} ${s}(t)$ defined on $t\in I \subset \mathbb{R}$ to the network state space, where $I$ is an interval: for each time $t \in I \subset \mathbb{R}$, ${s}(t)$ represents a random variable to the network state space:
\bq
\prob\left[{s}(t)  = {\bf S}\right] & \in & [0,1] \text{ for any state }{\bf S}\in \Sigma \nonumber \\
\sum_{{\bf S}\in \Sigma}\prob\left[{s}(t) = {\bf S}\right] & = & 1 
\eq

Notice that the random variables ${s}(t)$ are not independent, therefore $\prob\left[{s}(t)  = {\bf S},{s}(t')  = {\bf S}'\right] \neq \prob\left[{s}(t)  = {\bf S}\right]\prob\left[{s}(t')  = {\bf S}'\right]$. From now on, we define $\prob\left[{s}(t)  = {\bf S}\right]$ as {\sl instantaneous probabilities} (or first order probabilities). Since the instantaneous probabilities do not define the full stochastic process, all possible joint probabilities should also be defined.

In order to simplify the stochastic process, Markov property is imposed. It can be expressed in the following way: ``the conditional probabilities in the future, related to the present and the past, depend only on the present'' (see Supplementary Material, section 1.1 for the mathematical definition). A stochastic process with the Markov property is called a {\sl Markov process}.

Any Markov process can be defined by (see Van Kampen \cite{vanstochastic}, chapter IV):
\begin{enumerate}
\item an initial condition:
\be
\prob\left[{s}(0) = {\bf S}\right]\;; \forall{\bf S}\in \Sigma 
\ee
\item conditional probabilities (of a single condition):
\be
\prob\left[{s}(t)  = {\bf S}|{s}(t') = {\bf S}'\right]\; ; \forall {\bf S},{\bf S}' \in \Sigma \; ; \forall t',t \in I; t'<t
\ee
\end{enumerate}
  
Concerning time, two cases can be considered:
\begin{itemize}
\item Time is discrete: $t\in I=\{t_0,t_1,\cdots\}$. In that case, it can be shown\cite{shiryaev1996probability} that all possible conditional probabilities are function of {\sl transition probabilities}: $\prob\left[{s}(t_i)  = {\bf S}|{s}(t_{i-1}) = {\bf S}'\right]$. In that case, a Markov process is often named a Markov chain.
\item Time is continuous: $t\in I=[a,b]$. In that case, it can be shown\cite{vanstochastic} that all possible conditional probabilities are function of {\sl transition rates}: $\rho_{({\bf S}'\rightarrow {\bf S})}(t) \in [0,\infty[$ (see Supplementary Material, section 1.1 for the definition of transition rates).
\end{itemize}

Notice that a discrete time Markov process can be derived from continuous time Markov process, called {\sl Jump Process}, with the following transition probabilities:
\be
\prob_{{\bf S}\rightarrow {\bf S'}} \equiv \frac{\rho_{{\bf S}\rightarrow {\bf S'}}}{\sum_{{\bf S}''} \rho_{{\bf S}\rightarrow{\bf S}''}} \nonumber
\ee

If the transition probabilities or transition rates are time independent, the Markov process is called a {\sl time independent Markov process}. In BKMC, only this case will be considered. 
For a time independent Markov process, the {\sl transition graph} (often called Boolean state graph in the Boolean framework) can be defined as follows: a transition graph is a graph in $\Sigma$, with an edge between ${\bf S}$ and ${\bf S}'$ if and only if $\rho_{{\bf S}\rightarrow{\bf S}'}>0$ (or $\prob\left[{s}(t_i)  = {\bf S}|{s}(t_{i-1}) = {\bf S}'\right]>0$ if time is discrete). 
\subsubsection*{Asynchronous Boolean dynamics as a discrete time Markov process}
  
Asynchronous Boolean dynamics\cite{thomas1991regulatory} is widely used in Boolean modeling. It can be easily interpreted as discrete time Markov process\cite{chaves2005robustness,chaouiya2007petri} as shown below.

In the case of asynchronous Boolean dynamics, the system is given by $n$ nodes (or agents), with a set of directed arrows linking these nodes and defining a network. For each node $i$, a Boolean {\sl logic} $B_i({\bf S})$ is specified that depends only on the nodes $j$ for which there exists an arrow from node $j$ to $i$ (e.g. $B_1=S_3 \text{ AND} \text{ NOT } S_4$, where $S_3$ and $S_4$ are the Boolean values of nodes 3 and 4 respectively). The notion of {\sl asynchronous transition} (AT) can be defined as a pair of network states $({\bf S},{\bf S}') \in \Sigma$, written  $({\bf S} \rightarrow {\bf S}')$ such that 
\bq S'_j & = & B_j({\bf S})\text{ for a given }j \nonumber \\
S'_i & = & S_i\text{ for }i\neq j 
\eq

To define a Markov process, the transition probabilities $\prob\left[{s}(t_i)={\bf S}| {s}(t_{i-1})={\bf S}'\right]$ can be defined such that: given two network states ${\bf S}$ and ${\bf S}'$, let $\gamma({\bf S})$ be the number of asynchronous transitions from ${\bf S}$ to all possible states ${\bf S}'$. Then 
\bq
\prob\left[{s}(t_i)={\bf S}'| {s}(t_{i-1}) = {\bf S}\right] & = &1/ \gamma({\bf S})\text{ if }({\bf S}\rightarrow{\bf S}')\text{ is an AT} \nonumber \\
\prob\left[{s}(t_i)={\bf S}'| {s}(t_{i-1}) = {\bf S}\right] & = & 0 \text{ if }({\bf S}\rightarrow{\bf S}')\text{ is not an AT}
\eq

In this formalism, asynchronous Boolean dynamics completely defines a discrete time Markov process when the initial condition is specified. Notice that here the transition probabilities are time independent, i.e.
$\prob\left[{s}(t_i)={\bf S}| {s}(t_{i-1})={\bf S}'\right]=\prob\left[{s}(t_{i+1})={\bf S}| {s}(t_i)={\bf S}'\right]$. Other definition of $\gamma({\bf S})$ can be used, defining another Markov process that have the same transition graph. Therefore,
the approaches mention above, that introduce time implicitly by adding probabilities to each transition of the transition graph, can be seen as a generalization of the definition of $\gamma({\bf S})$. 


\subsubsection*{Continuous time Markov process as a generalization of asynchronous Boolean Dynamics}

To transform the discrete time Markov process described above in a continuous time Markov process, transition probabilities should be replaced by transition rates $\rho_{({\bf S} \rightarrow {\bf S}')}$. In that case, conditional probabilities are computed by solving a {\sl master equation} (equation 2 in Supplementary Material, section 1.1). We present below the corresponding numerical algorithm ({\sl Kinetic Monte-Carlo} \cite{young1966monte}).

Because we want a generalization of asynchronous Boolean dynamics, transition rates $\rho_{({\bf S} \rightarrow {\bf S}')}$ are non-zero \underline{only if} ${\bf S}$ and ${\bf S}'$ differ by only one node. 
In that case, each Boolean logic $B_i({\bf S)}$ is replaced by two functions $R_i^\text{up/down}({\bf S})\in [0,\infty[$. The transition rates are defined as follow: if $i$ is the node that differs from ${\bf S}$ and ${\bf S}'$, 
\bq
\rho_{({\bf S} \rightarrow {\bf S}')} & = & R^\text{up}_i({\bf S})\text{ if }S_i=0 \nonumber \\
\rho_{({\bf S} \rightarrow {\bf S}')} & = & R^\text{down}_i({\bf S})\text{ if }S_i=1 \label{eq:R2rho}
\eq

$R^\text{up}_i$ corresponds to the activation rate of node $i$, $R^\text{down}_i$ corresponds to the inhibition rate of node $i$. 
Therefore, the continuous Markov process is completely defined by all these $R^\text{up/down}$ and an initial condition.


\subsubsection*{Asymptotic behavior of continuous time Markov process}

In the case of continuous time Markov process, instantaneous probabilities always converge to a stationary distribution (see Supplementary Material, corollary 2, section 1.2).  A {\sl stationary distribution} of a given Markov process corresponds to the set of instantaneous probabilities of a stationary Markov process which has the same transition probabilities (or transition rates) of the given discrete (or continuous) time Markov process. A {\sl stationary Markov process} has the following property: for every joint probability $\prob\left[{s}(t_1)={\bf S}^{(1)},{s}(t_2)={\bf S}^{(2)},\ldots\right]$ and $\forall \tau$,
\be
\prob\left[{s}(t_1)={\bf S}^{(1)},{s}(t_2)={\bf S}^{(1)},\ldots\right]=\prob\left[{s}(t_1+\tau)={\bf S}^{(1)},{s}(t_2+\tau)={\bf S}^{(1)},\ldots\right]
\ee

Notice that instantaneous probabilities $\prob\left[{s}(t)={\bf S}\right]$ of a stationary stochastic process are time independent. 

The asymptotic behavior of a continuous time Markov process can be detailed by using the concept of indecomposable stationary distributions: {\sl indecomposable stationary distributions} are stationary distributions that cannot be expressed as linear combination of different stationary distributions. Notice that a linear combination of stationary distributions is also a stationary distribution, up to a constant. This comes from the fact that instantaneous probabilities are solutions of a master equation, which is linear (see Supplementary Material, equation 2, section 1.1). Therefore, a complete description of the asymptotic behavior is given by the linear combination of indecomposable stationary distributions, to which the Markov process converges. 


\subsubsection*{Oscillations and cycles}

In order to describe a periodic behavior, the notion of cycle and oscillation for a continuous time Markov process is defined precisely. 

A {\sl cycle} is a loop in the transition graph. 
 This is a topological characterization that does not depend on the exact value of transition rates. It can be shown that a cycle with no outcoming edge corresponds to an indecomposable stationary distribution (see Supplementary Material, corollary 1, section 1.2).

The question is then to link the notion of cycle to that of periodic behavior of instantaneous probabilities. Instantaneous probabilities cannot be perfectly periodic; at most, they have a damped ocscillating behavior (see Supplementary Material, section 1.3). Let us define formally a {\sl damped oscillatory} Markov process as a continuous time process that has at least one instantaneous probability with an infinite number of extrema.

According to theorems described in Supplementary Material (theorems 6-8  and Corollary 3, section 1.3), a necessary condition for having damped oscillation is that the transition matrix (see Supplementary Material, equation 4, section 1.1) has at least one non-real eigenvalue. In that case, there always exists an initial condition that produces damped oscillations. For the transition matrix to have a non-real eigenvalue, a Markov process needs to have a cycle. However, the reverse is not true. In the toy model of single cycle, presented in the section of examples, non-real eigenvalues may or may not exist, according to different sets of transition rates, although the transition graph remains the same.

\subsubsection*{BKMC: Kinetic Monte-Carlo (Gillespie algorithm) applied to continuous time asynchronous Boolean Dynamics}

It has been previously stated that a continuous time Markov process is completely defined by its initial condition and its transition rates. For computing any conditional probability (and any joint probability), a set of linear differential equations has to be solved (the {\sl master equation}). Theoretically, the master equation can be 
solved exactly by computing the exponential of the transition matrix (see Supplementary Material, equation 5, section 1.1). However, because the size of this transition matrix is $2^n\times 2^n$, practical computation soon becomes impossible if $n$ is large. To remedy this problem, it is possible to use a simulation algorithm that samples the probability space by computing time trajectories in the network state space.

The Kinetic Monte-Carlo\cite{young1966monte} (or Gillespie algorithm\cite{gillespie1976general}) is a simple algorithm for exploring the probability space of a Markov process defined by a set of transition rates. In fact, it can be understood as a formal definition of a continuous time Markov process. This algorithm produces a set of {\sl realizations} or {\sl stochastic trajectories} of the Markov process, given a set of uniform random numbers in $[0,1[$. By definition, a trajectory $\hat{\bf S}(t)$ is a function from a time window $[0,t_\text{max}]$ to $\Sigma$. The set of {\sl realizations} or {\sl stochastic trajectories} represents the given Markov process in the sense that these trajectories can be used to computed probabilities. Practically, a finite set of these trajectories is produced, then probabilities are estimated from this finite set (as described below). The algorithm is based on an iterative step: from a state ${\bf S}$ at time $t_0$ (given two uniform random numbers), it produces a transition time $\delta t$ and a new state ${\bf S'}$, with the following interpretation: the trajectory $\hat{\bf S}(t)$ is such that $\hat{\bf S}(t)={\bf S}$ for $t \in [t_0,t_0+\delta t[$ and $\hat{\bf S}(t_0+\delta t)={\bf S}'$. Iteration of this step is done until a specified maximum time is reached. The initial state of each trajectory is based on the (probabilistic) initial condition, that needs also to be specified. 

The exact iterative step is the following, given ${\bf S}$ and two uniform random number $u,u'\in [0,1[$:
\begin{enumerate}
\item Compute the total rate of possible transitions for leaving ${\bf S}$ state: \\
$\rho_\text{tot}\equiv \sum_{{\bf S'}}\rho_{({\bf S}\rightarrow {\bf S'})}$. 
\item Compute the time of the transition: $\delta t \equiv -\log(u)/\rho_\text{tot}$ 
\item Order the possible new states ${\bf S}'^{(j)}, j=1 \dots$ and their respective transition rates $\rho^{(j)}=\rho_{({\bf S}\rightarrow {\bf S'^{(j)}})}$.
\item Compute the new state ${\bf S}'^{(k)}$ such that $\sum_{j=0}^{k-1}\rho_j<(u'\rho_\text{tot})\leq\sum_{j=0}^k\rho_j$ (by convention, $\rho^{(0)}=0$).
\end{enumerate}

The application of this algorithm to continuous time Markov process in network state space will be referred to as {\sl Boolean Kinetic Monte-Carlo} or BKMC.

\subsection*{Practical use of BKMC, through MaBoSS tool}

Biological data are summarized into an influence network with logical rules associated to each node of the network. The value of one node depends on the value of the input nodes. For BKMC, another layer of information is provided when compared to the standard definition of Boolean models: transition rates are provided for all nodes, specifying the rates at which the node turns on and off based on their logic for both the on and off rules. This refinement conserves the simplicity of Boolean description but allows to reproduce the observed biological dynamics. The parameters do not need to be exact as in nonlinear ordinary differential equation models but they can be used to illustrate the relative speed of reactions. For that purpose, we developed a software tool, MaBoSS, that applies BKMC algorithm. MaBoSS stands for Markov Boolean Stochastic Simulator. Practically, MaBoSS needs two input files: one describing the network and its transition rates, one describing the parameters controlling the different estimates described in the Methods section. Source code, reference card and examples are available on the web: https://maboss.curie.fr.

\subsubsection*{Transition rates with MaBoSS language: modeling biological processes}

MaBoSS defines transition rates $\rho_{({\bf S} \rightarrow {\bf S}')}$ by the functions $R^\text{up/down}_j({\bf S})$ (see equation \ref{eq:R2rho}). The format of these functions is very flexible. It includes all Boolean operators (AND, OR, NOT, XOR), arithmetic operators (+,-,* /), external variables, node variables, comparison operators and the conditional operator (?:). Examples of the use of the language are given below to illustrate three different cases: different speeds for different inputs, buffering effect and the translation of discrete variables (with more than the 2 values, 0 and 1) in MaBoSS.    
\begin{itemize}
\item Modeling different speeds of incoming influences:
suppose that C is activated by A and B, but that B can activate C faster than A. In this case, we write:

\begin{verbatim}
node C {
rate_up= B ? $kb : (A ? $ka : 0.0);
rate_down= (A | B ) ? 0.0 : 1.0}
\end{verbatim}

When C is off (equal to 0), C is activated by B at a speed kb. If B is absent, then C is activated by A at a speed \$ka. If both are absent, C is not activated. Note that if both A and B are present, because of the way the logic is written in this particular case, C is activated at a speed \$kb. 
When C is on (equal to 1), C is inactivated at a rate equal to 1 if A and B are both absent. 

To implement the synergetic effect of A and B, \emph{i.e.} when both A and B are on, C activates at a rate \$kab, then we can write:
\begin{verbatim}
node C {
rate_up= (A & !B ? $ka : 0)+(B & !A ? $kb : 0) + (A & B ? $kab : 0.0);
rate_down= (A | B ) ? 0.0 : 1.0}
\end{verbatim}

\item Modeling buffering effect:
suppose that B is activated by A, but that B can remain active a long time after A has shut down. For that, it is enough to define different speeds of activation and inhibition:

\begin{verbatim}
node B {
rate_up= A ? 2.0 : 0.0;
rate_down= A ? 0.0 : 0.001;}
\end{verbatim}

B is activated by A at a rate equal to 2. When A is turned off, B is inactivated more slowly at a rate equal to 0.001.
\item Modeling more than two discrete states for a given node:
Suppose that B is activated by A, but if the activity of A is maintained, B can reach a second level. For this, we define a second node B\_h (for ``B high'') with the following rules:

\begin{verbatim}
node B {
rate_up= A ? 1.0 : 0.0;
rate_down= (A | B_h) ? 0.0 : 1.0;}

node B_h {
rate_up= (A & B) ? 1.0 : 0.0;
rate_down= (A) ? 0.0 : 1.0;}
\end{verbatim}

In this example, B is separated in two variables: B which corresponds to the first level of B and B\_h which corresponds to the higher level of B. 
B is activated by A at a rate equal to 1. If A disappears before B has reached its second level B\_h then B is turned off at a rate equal to 1. If A is maintained and B is active, then B\_h is activated at a rate equal to 1. When A is turned off, B\_h is inactivated at a rate equal to 1. 
\end{itemize}


\subsubsection*{Simulation input parameters in MaBoSS}

To simulate a process in MaBoSS, a set of parameters need to be adjusted (see simulation parameters in reference card). MaBoSS assigns default values, however, they need to be tuned for each model to achieve optimal performances: best balance between the convergence of estimates and the computation time. Therefore, several simulations should be run with different set of parameters for best tuning. 

\begin{itemize}
\item Internal nodes: \emph{node.is\_internal}
As explained in Methods (``Initial conditions and outputs''), internal nodes correspond to species that are not measured explicitly. Practically, the higher the number of internal nodes, the better the convergence of the BKMC algorithm. 

\item Time window for probabilities: \emph{timetick}

This parameter is used to compute estimates of network state probabilities (see ``Network state probabilities on time window'' in Methods). A time window can be set as the minimum time needed for nodes to change their states. This parameter also controls the convergence of probability estimates. The larger the time window parameter, the better the convergence of probability estimates. With practice, the tradeoff between timetick parameter value and the convergence speed will be defined. 

\item Maximum time: \emph{max\_time}

MaBoSS produces trajectories for a predefined amount of time set by the parameter: max\_time. This maximum time needs to be specified. 
If the time of the biological process is known, then the maximum time parameter can be explicitly set. 
If the time of the biological process is not known, then there exists a more empirical way to set the maximum time. It is advised to choose a maximum time parameter that is slightly bigger than the inverse of the smallest transition rate. 

Note that the computing time in MaBoSS is proportional to this maximum time. Moreover, the choice of the maximum time impacts the stationary distribution estimates (see below): a longer maximum time increases the quality of these estimates.

\item Number of trajectories: \emph{sample\_count}

This parameter directly controls the quality of BKMC estimation algorithm. Practically, the convergence of the estimates increases as the number of trajectories is increased.

\item Number of trajectories (\emph{statdist\_traj\_count}) and similarity threshold (\emph{statsdist\_cluster\_threshold}) for stationary distribution estimates

The \emph{statdist\_traj\_count} parameter corresponds to a subset of trajectories use only for stationary distribution estimates (the \emph{statdist\_traj\_count} first trajectories are chosen by the algorithm). To avoid explosion of computing time, this parameter needs to be lower than the number of trajectories (rather than equal to).

The \emph{statsdist\_cluster\_threshold} parameter corresponds to the threshold for constructing the clusters of stationary distribution estimates. Ideally, it should be set to a high value (close to 1). However, if the threshold is too high then the clustering algorithm might not be efficient. 

For optimal results, the identification of the full set of indecomposable stationary distributions should be done in the following way:
\begin{itemize}
\item Run a simulation with an initial condition and maximum time, with a reasonable similarity threshold (around 0.8) and a number of trajectories (around 1000). As a response, MaBoSS provides a set of indecomposable stationary distributions (it corresponds to the stationary distributions associated to each cluster).
\item Select the states with non-zero probability in the set of indecomposable stationary distributions and set them as initial conditions, increase the maximum time (\emph{max\_time}), the number of trajectories (\emph{statdist\_traj\_count}) and/or the similarity threshold (\emph{statsdist\_cluster\_threshold}).
\item Run the simulations for these new parameters.
\item If new indecomposable stationary distributions appear, start again the two previous steps. Stop when the indecomposable stationary distributions remain stable with respect to the simulation parameters, \emph{i.e.} after several rounds. 
\end{itemize}
 
\end{itemize} 

\subsubsection*{Comparison with biological data}

The relationship between a model and experimental data is strongly dependent on the type of model and the question that needs to be solved. 

Because MaBoSS is based on Boolean modeling, the biological data need to be discretized. Each node of the model should represent discrete levels of the respective species (mRNA, protein, protein complex, etc.). It is possible to have more than two discrete levels in a model, as shown in the example ``Modeling more than two discrete states for a given node''.

The transitions rates are positive numbers that should be introduced in a model; it is possible to extract them from experimental data, using the following property: the rate of a given transition is the inverse of the mean time, for this transition to happen. It should be noticed than BKMC is an algorithm based on a linear equation (Supplementary Material, equation 2, section 1.1); therefore, small variations of transition rates won't affect qualitative behavior of a model.

The basic outputs of BKMC algorithm are network state probabilities over time. These can be interpreted in terms of a cell population, once experimental data are discretized. The asymptotic behavior of a model, represented by a linear combination of indecomposable stationary distributions, can be interpreted as a combination of cell sub-populations. More precisely, consider an indecomposable stationary distribution and the associated set of network states with non-zero probability. A cell in such a network state can only evolve in other network states with non-zero probability, within the same indecomposable stationary distribution (Supplementary Material, corollary 1, section 1.2 and by using the definition of strongly connected component with no outcoming edge). Therefore, the set of network states with non-zero probability can be interpreted as a sub-population whose cells evolve only within the sub-population.

\subsection*{Examples}

Two models using BKMC applied to Boolean networks are given as examples. The first one is a toy model, illustrating the dynamics of a single cycle. The second one uses a published Boolean model of p53-Mdm2 response to DNA damage. Note that MaBoSS has been used for these two examples, but Markov process can be computed directly, without our BKMC algorithm because the model is small enough (by computing exponential of transition matrix, see Supplementary Material, section 1.1), as proposed in \cite{teraguchi2011stochastic}. 
BKMC is best suited for larger networks, when the network state space is too large to be computed with standard existing tools ($ >\sim 2^{10}$). These examples were chosen for their simplicity, and because they illustrate how global characterizations (entropy and transition entropy, see ``Entropies'' in Methods) can be used.

For the purpose of this article, we built the transition graphs for both examples (with GINsim \cite{gonzalez2006ginsim}) in order to help the reasoning. However, BKMC algorithm does not construct the transition graph explicitly. 

\subsubsection*{Toy model of a single cycle}
 
We consider three species, A, B and C, where A is activated by C and inhibited by B, B is activated by A and C is activated by A or B (Figure \ref{fig:model_toy}). 

\begin{figure}[!ht]
\includegraphics[width=17cm]{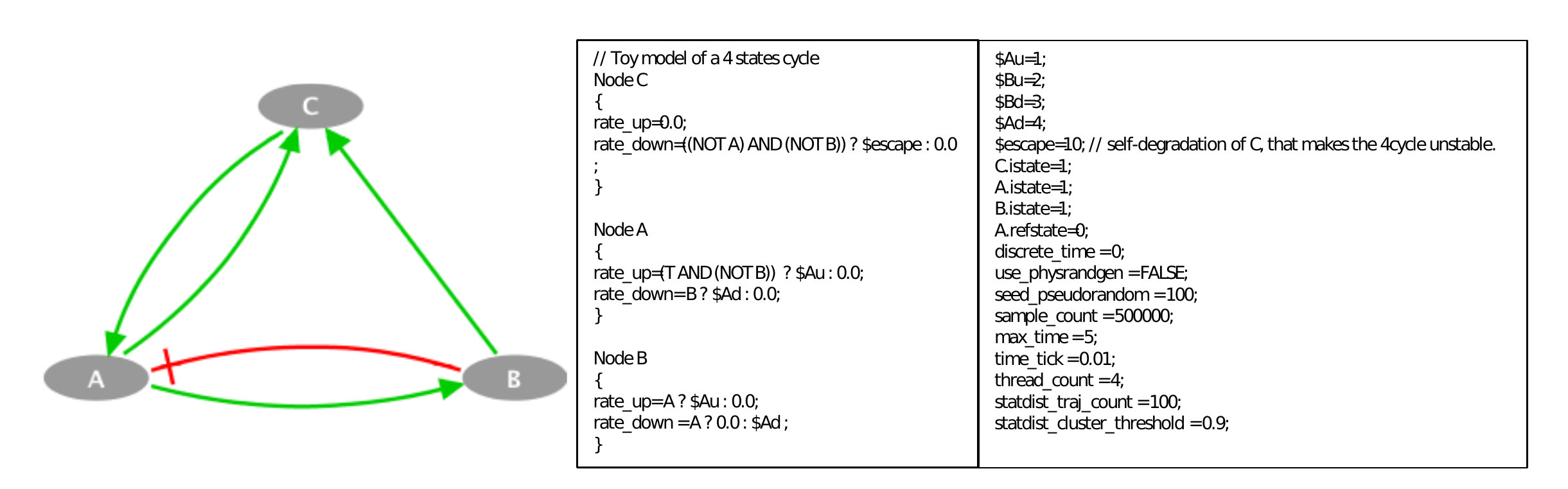}
\caption{Toy model of a single cycle. (A) Influence network. (B) Logic and transition rates of the model. (C) Simulation parameters}
\label{fig:model_toy}
\end{figure}

The model is defined within the language of MaBoSS (defined in the web page, https://maboss.curie.fr). The associated transition graph is shown in figure \ref{fig:state_toy}. 
In this model, the only stationary distribution is the fixed point [ABC]=[000]. We studied two cases: when the rate of the transition from state [001] to state [000] is either fast or slow (inactivation of C). We will refer to this transition rate as the {\sl escape rate}.
 
\begin{figure}[!ht]
\includegraphics[width=5cm]{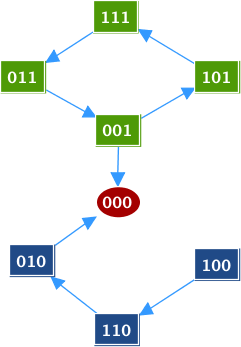}
\caption{Transition graph for the toy model. The node state should be read as [ABC] = [***]. [ABC]=[100] corresponds to a state in which only A is active. The nodes in green belong to a cycle, the node in red is the fixed point and the other state nodes are in blue.}
\label{fig:state_toy}
\end{figure}

In the first case, when the escape rate is fast, we set the parameter for the transition to a high value (rate\_up = 10). In figure \ref{fig:FourCycleFast_traj}, we notice that the probability that [ABC] is equal to [000] converges to 1. We can conclude that [ABC]=[000] is a fixed point. In addition, the entropy and the transition entropy converge to zero. With BKMC, these properties correspond to a signature of a fixed point. The peak in the entropy (between times 0 and 0.6) corresponds to a set of transiently activated states before reaching the fixed point. 

\begin{figure}[!ht]
\includegraphics[scale=0.6]{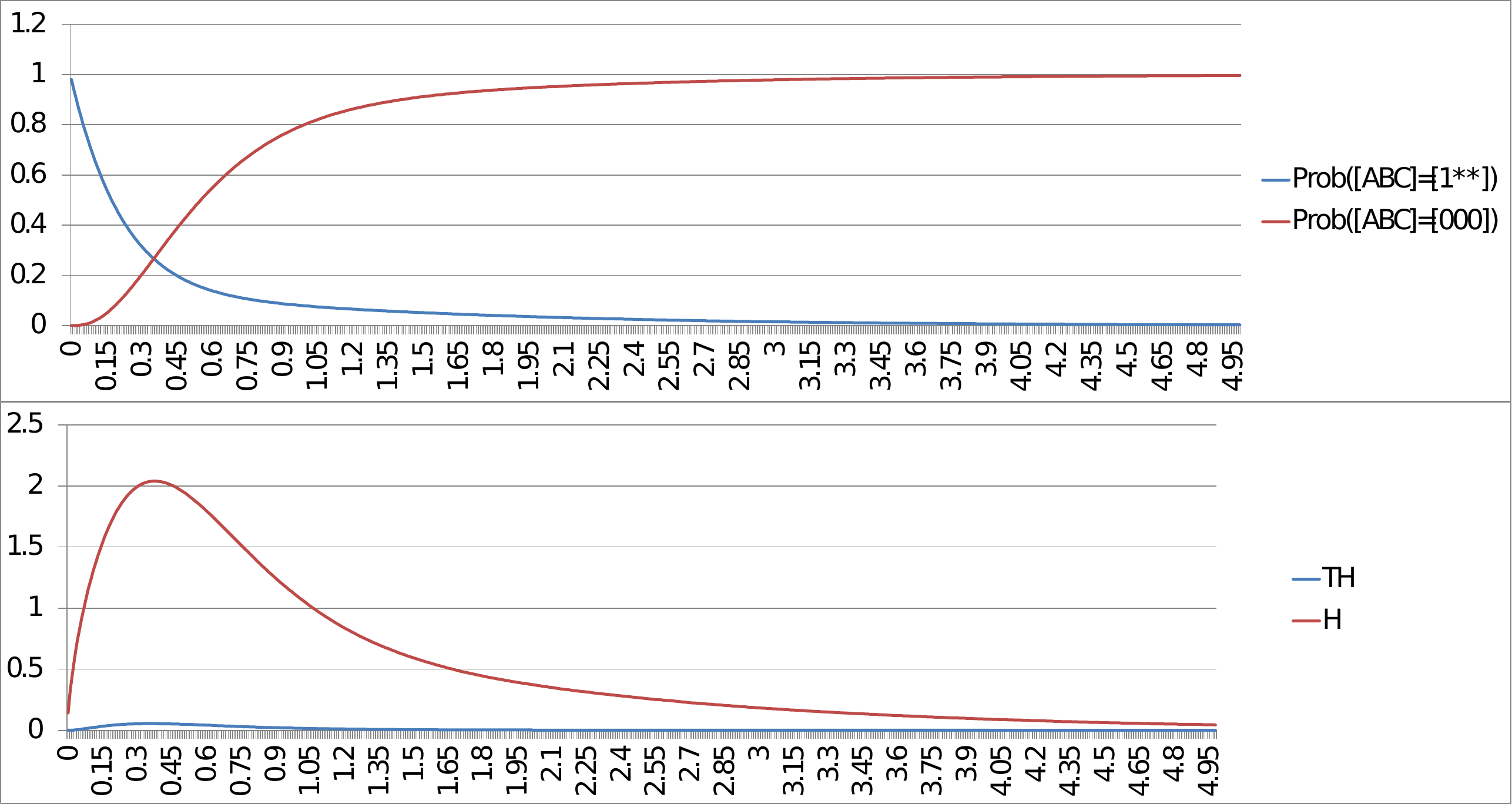}
\caption{BKMC algorithm applied to the toy model, with a fast escape rate. Time trajectory of probabilities ([ABC]=[000] and [ABC]=[1**] where * can be either 0 or 1), the entropy ($H$) and the transition entropy ($TH$) are plotted. Because the probability of [ABC]=[000] converges to 1, [ABC]=[000] is a fixed point. The asymptotic behavior of both the entropy and the transition entropy is also the signature of a fixed point.}
\label{fig:FourCycleFast_traj}
\end{figure}

In the second case, when the escape rate is slow, we set the parameter for the transition to a low value (rate\_down = $10^{-5}$). The model seems to show another stationary distribution. As illustrated in figure \ref{fig:FourCycleSlow_traj}, the transition entropy is and remains close to zero but the entropy does not converge to zero, which is the signature of a cyclic stationary distribution (see ``Entropies'' in Methods). This corresponds to the cycle [111] $\rightarrow$ [011] $\rightarrow$ [001] $\rightarrow$ [101] in the transition graph (\ref{fig:state_toy}). However, as seen in the transition graph, one state in the cycle will eventually lead the fixed point (through the transition [001] $\rightarrow$ [000] in figure \ref{fig:state_toy}). Therefore, if the temporal evolution is plotted on a larger time scale (Figure \ref{fig:FourCycleSlow_traj_long}), it looks similar to the case of fast escape rate. This case can be anticipated. Indeed, the value of the transition entropy of figure \ref{fig:FourCycleSlow_traj} is not exactly zero, but $10^{-4}$. Therefore, the cyclic behavior is not stable. We can conclude on stable cyclic behaviors only when the transition entropy is exactly zero. 

By considering the spectrum of the transition matrix (see Supplementary Material, section 1.1 and proof of theorem 4), it can be proven that the model with a slow escape rate is a damped oscillatory process whereas the model with a large escape rate is not a damped oscillatory process. As mentioned previously, a cycle in the transition graph may or may not lead to an oscillatory behavior. Moreover, if the transition entropy seems to converge to a small value on a small time scale, and the entropy does not, this behavior illustrates the case of a transient cycle in the transition graph. 
 
\begin{figure}[!ht]
\includegraphics[scale=0.6]{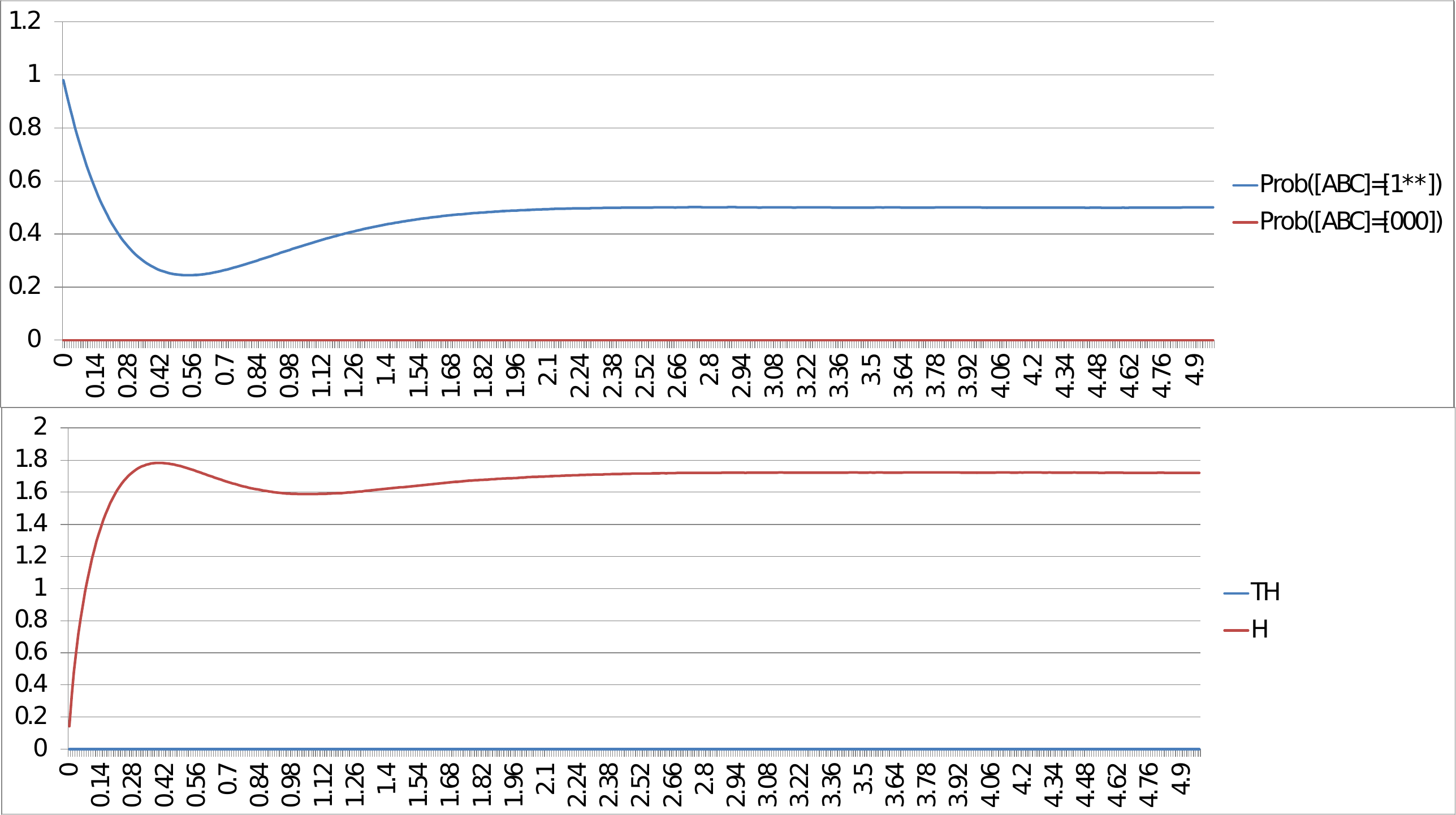}
\caption{BKMC algorithm applied to the toy model, with a slow escape rate. Time trajectory of probabilities ([ABC]=[000] and [ABC]=[1**]), the entropy ($H$) and the transition entropy ($TH$) are plotted. The asymptotic behavior of both the entropy and the transition entropy seems to be the signature of a cycle.}
\label{fig:FourCycleSlow_traj}
\end{figure}

\begin{figure}[!ht]
\includegraphics[scale=0.6]{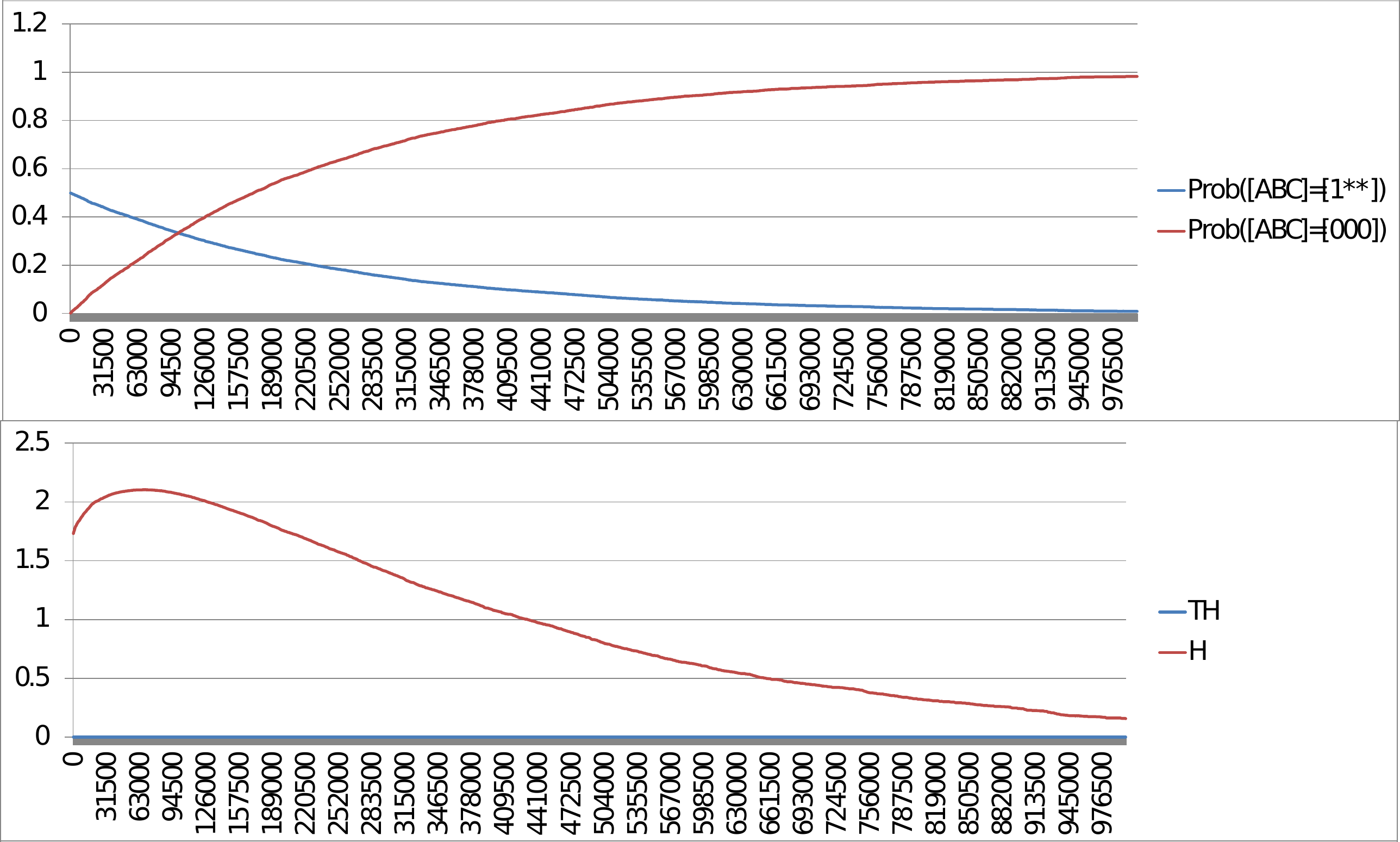}
\caption{BKMC algorithm applied to the toy model, with a slow escape rate, plotted on a larger time scale. Time trajectory of probabilities ([ABC]=[000] and [ABC]=[1**]), the entropy ($H$) and the transition entropy ($TH$) are plotted. On a large time scale, the asymptotic behavior of both the entropy and the transition entropy is similar to the case of the fast escape rate (figure \ref{fig:FourCycleFast_traj}).}
\label{fig:FourCycleSlow_traj_long}
\end{figure}

\subsubsection*{p53-Mdm2 signaling}
We consider a model of p53 response to DNA damage \cite{p53Model}.
p53 interacts with Mdm2, which appears in two forms, cytoplasmic and nuclear. On one hand, p53 upregulates the level of cytoplasmic Mdm2 which is then transported into the nucleus and inhibits the export of nuclear Mdm2. On the other hand, Mdm2 facilitates the degradation of p53 through ubiquitination. In the model, stress regulates the level of DNA damage, which in turn participates in the degradation process of Mdm2. p53 inhibits the DNA damage signal by promoting DNA repair. Here, stress is not shown explicitly (Figure \ref{fig:model_damp53}). 

\begin{figure}[!ht]
\includegraphics[scale=0.5]{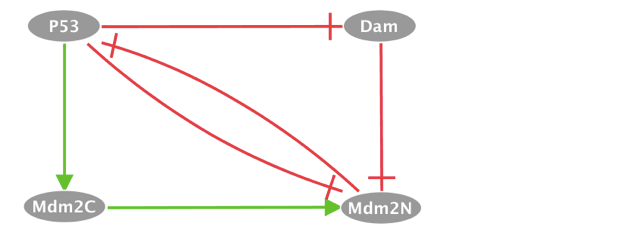}
\caption{Boolean model of p53 response to DNA damage}
\label{fig:model_damp53}
\end{figure}

The model is defined within the language of MaBoSS, with two levels of p53 as it is done in Abou-Jaoud\'e \emph{et al.}  \cite{p53Model}. The model is implemented in MaBoSS (provided in the web page (https://maboss.curie.fr) along with the simulation parameters). The associated transition graph is given in figure \ref{fig:state_damp53}. It shows the existence of two cycles and of a fixed point [p53 Mdm2C Mdm2N Dam] = [0010]. 

\begin{figure}[!ht]
\includegraphics[scale=0.5]{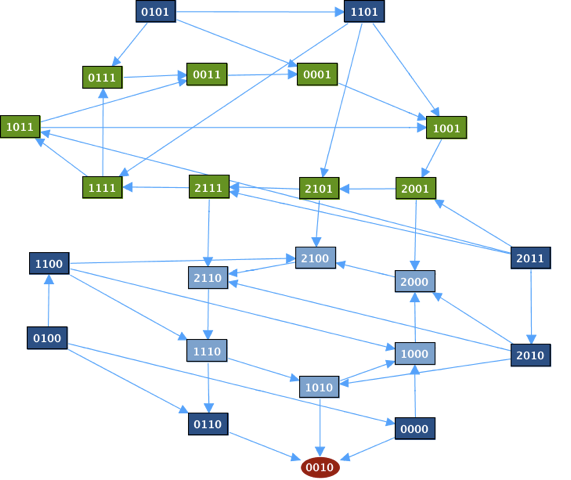}
\caption{Transition graph of the p53 model.The node states should be read as [p53 Mdm2C Mdm2N Dam] = [****] (where * can be either 0 or 1). For instance, [p53 Mdm2C Mdm2N Dam]=[1000] corresponds to a state in which only p53 (at its level 1) is active. The nodes in green and the nodes in light blue belong to two cycles, the node in red is the fixed point and the other state nodes are in dark blue.}
\label{fig:state_damp53}
\end{figure}

In order to represent the activity of p53, time evolution of its expectation value is shown in figure \ref{fig:prob_damp53}, with the initial condition: [p53 Mdm2C Mdm2N Dam] = [0*11]. The activation of p53 seems to be transient. 


\begin{figure}[!ht]
\includegraphics[scale=0.5]{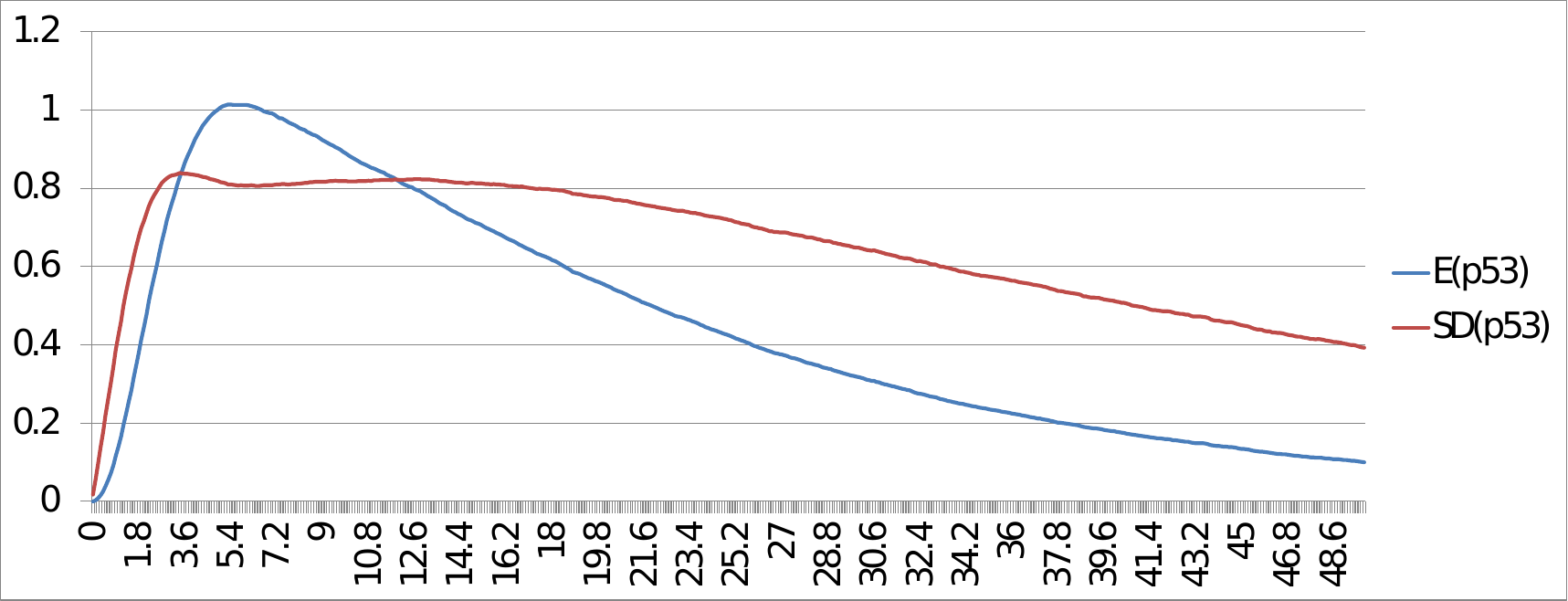}
\caption{Time trajectories of p53 expectation value and standard deviation}
\label{fig:prob_damp53}
\end{figure}

The qualitative results obtained with MaBoSS are similar to those of Abou-Jaoud\'e and colleagues. However, at the level of cell population, some discrepancies appear: in figure \ref{fig:prob_damp53}, no damped oscillations can be seen as opposed to figure 8 of their article. The reason is that in their computations, noise imposed on time is defined by a square distribution on a limited time frame, whereas in BKMC, Markovian hypotheses imply that the noise distribution is more spread out from 0 to infinity. The consequence is that synchronization is lost very fast. Damped oscillations could be observed with BKMC with a particular set of parameters: fast activation of p53 and slow degradation of p53 (results not shown). 

With MaBoSS, we clearly interpret the system as a population and not as a single cell. In addition, we can simulate different contexts, presented in the initial article as different models, within one model that used different parameters to account for these contexts (see web page https://maboss.curie.fr for more details).

\begin{figure}[!ht]
\includegraphics[scale=0.5]{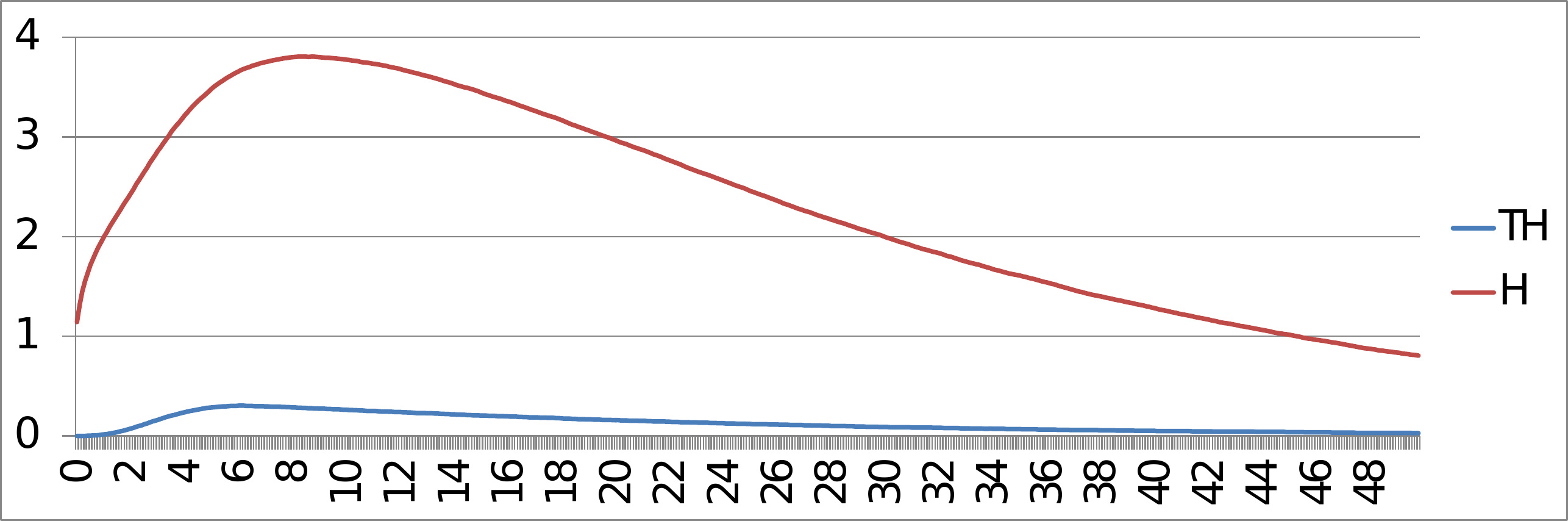}
\caption{Time trajectories of the entropy ($H$) and the transition entropy ($TH$)}
\label{fig:H_damp53}
\end{figure}

Note that the existence of transient cycles, as shown in the toy model, can be deduced from the time trajectory of the entropy that is significantly higher than the time trajectory of the transition entropy (which is non zero therefore the transient cycles are not stable) (Figure \ref{fig:H_damp53}). An other indirect effect of transient cycles is the flat part in p53 standard deviation, in figure \ref{fig:prob_damp53}.

\section*{Conclusions}

In this work, we present a new algorithm, Boolean Kinetic Monte-Carlo or BKMC, applicable to dynamical simulation of signaling networks based on continuous time in the Boolean framework. BKMC algorithm is a natural generalization of asynchronous Boolean dynamics \cite{thomas1991regulatory}, with time trajectories that can be interpreted in terms of biological time. The variables of the Boolean model are biological species and the parameters are rates of activation or inhibition of these species. These parameters can be deduced from experimental data.
We applied this algorithm to two different models: a toy model that illustrates a simple cyclic behavior, and a published model of p53 response to DNA damage.

This algorithm is provided within a freely available software, MaBoSS, that can run BKMC algorithm on networks up to 64 nodes, in the present version. The construction of a model uses a specific grammar that allows to introduce logical rules and rates of node activation/inhibition in a flexible manner. This software provides global and semi-global outputs of the model dynamics that can be interpreted as signatures of the dynamical behaviors. These interpretations become particularly useful when the network state space is too large to be handled. The convergence of BKMC algorithm can be controlled by tuning different parameters: maximum time of the simulation, number of trajectories, length of a time window on which the average of probabilities is performed, and the threshold for the definition of stationary distributions clusters.

The next step is to apply BKMC algorithm with MaBoSS on large signaling networks, e.g. EGFR pathway, cell cycle signaling, apoptosis pathway, etc. The translation of existing Boolean models in MaBoSS is straightforward but requires the addition of transition rates. In these future works, we expect to illustrate the advantage of BKMC on other simulation algorithms. 
Moreover, in future developments of MaBoSS, we plan to introduce methods for sensitivity analyses and to refine approximation methods used in BKMC, and generalize Markov property. 

\section*{Methods}

BKMC generates stochastic trajectories. Here, we describe how we use and interpret them.

\subsubsection*{Network state probabilities on time window} \label{sect:prob_time_window}

To relate continuous time probabilities to real processes, an observable time window $\Delta t$ is defined. A discrete time ($\tau \in \mathbb{N}$) stochastic process ${s}(\tau)$ (that is {\bf not} necessary Markovian) can be extracted from the continuous time Markov process:
\be
\prob\left[{s}(\tau)={\bf S}\right] \equiv \frac{1}{\Delta t} \int_{\tau\Delta t}^{(\tau+1)\Delta t} dt\; \prob\left[{s}(t)={\bf S}\right] 
\ee 

BKMC is used for estimating $\prob\left[{s}(\tau)={\bf S}\right]$ as follows:
\begin{enumerate}
\item For each trajectory $j$, compute the time for which the system is in state ${\bf S}$, in the window $[\tau\Delta t,(\tau+1)\Delta t]$. Divide this time by $\Delta t$. Obtain an estimate of $\prob\left[{s}(\tau)={\bf S}\right]$ for trajectory $j$, \emph{i.e.} $\hat{\prob}_j\left[{s}(\tau)={\bf S}\right]$.
\item Compute the average over $j$ of all $\hat{\prob}_j\left[{s}(\tau)={\bf S}\right]$ to obtain $\hat{\prob}\left[{s}(\tau)={\bf S}\right]$. Compute the error of this average ($\sqrt{\text{Var}(\hat{\prob}\left[{s}(\tau)={\bf S}\right])/\text{\# trajectories}}$).
\end{enumerate}
\subsubsection*{Entropies\label{sect:entropies}}

Once $\prob\left[{s}(\tau)={\bf S}\right]$ is computed, the entropy $H(\tau)$ can be estimated
\be
H(\tau)=-\sum_{{\bf S}}\log_2\left(\prob\left[{s}(\tau)={\bf S}\right]\right) \prob\left[{s}(\tau)={\bf S}\right]
\ee
The entropy measures the disorder of the system. A maximum entropy means that all states have the same probability, a zero entropy means that one of the states has a probability of one. The estimation of the entropy can be seen as a global characterization (by a single real number) of a full probability distribution. The choice of $\log_2$ allows to interpret $H(\tau)$ in an easier manner: $2^{H(\tau)}$ is an estimate of the number of states that have a non-negligible probability in the time window $[\tau\Delta t,(\tau +1)\Delta t]$. A more computer-like interpretation of $H(\tau)$ is the number of bits that is necessary for describing states of non-negligible probability. 

The {\sl Transition Entropy} $TH$ is a finer measure that characterizes the system at the single trajectory level; it can be computed in the following way: for each state ${\bf S}$, there exists a set of possible transitions ${\bf S}\rightarrow {\bf S}'$. For each of these transitions, a probability is associated: 
\be
\prob_{{\bf S}\rightarrow {\bf S'}} \equiv \frac{\rho_{{\bf S}\rightarrow {\bf S'}}}{\sum_{{\bf S}'} \rho_{{\bf S}\rightarrow{\bf S}'}} \label{eq:ProbTrans}.
\ee

By convention, $\prob_{{\bf S}\rightarrow {\bf S'}}=0$ if there is no transition from ${\bf S}$ to any other state.

Therefore, the transition entropy $TH$ can be associated to each state ${\bf S}$:
\be
TH({\bf S})=-\sum_{{\bf S}'} \log_2(\prob_{{\bf S}\rightarrow {\bf S'}})\prob_{{\bf S}\rightarrow {\bf S'}} \label{eq:TransEnt}
\ee
Similarly, $TH({\bf S})=0$ if there is no transition from ${\bf S}$ to any other state.
The {\sl discrete time transition entropy} $TH(\tau)$ is defined as: 
\be
TH(\tau)=\sum_{{\bf S}}\prob\left[{s}(\tau)={\bf S}\right]TH({\bf S}) \nonumber
\ee

This transition entropy is estimated in the following way:
\begin{enumerate}
\item For each trajectory $j$, compute the set ($\Phi$) of visited states (${\bf S}$) in time window $[\tau\Delta t,(\tau+1)\Delta t]$ and their respective duration ($\mu_{\bf S})$. The estimated transition entropy is 
\be 
\hat{TH(\tau)}_j=\sum_{{\bf S}\in \Phi}TH({\bf S})\frac{ \mu_{\bf S}}{\Delta t}
\ee
\item Compute the average over $j$ of all $\hat{TH(\tau)}_j$ to obtain $\hat{TH(\tau)}$. Compute the error of that average ($\sqrt{\text{Var}(\hat{TH(\tau)})/\text{\# trajectories}}$).
\end{enumerate}
This transition entropy is a way to measure how deterministic the dynamics is. If the transition entropy is always zero, the system can only make a transition to a given state (although time of transitions remains stochastic). 

If probability distributions on time window tend to constant values (or tend to a stationary distribution), the entropy and the transition entropy can help characterize this stationary distribution such that:
\begin{itemize}
\item A fixed point has zero entropy and zero transition entropy,
\item A cyclic stationary distribution has non-zero entropy and zero transition entropy.
\end{itemize}

Entropy and transition entropy can be considered as ``global characterization'' of time evolution model: for a given time window, they always consist of two real numbers, whatever the size of the network.
\subsubsection*{Hamming distance distribution}

The {\sl Hamming Distance} between two states ${\bf S}$ and ${\bf S}'$ is the number of nodes that have different node states between ${\bf S}$ and ${\bf S}'$: 
\be
HD({\bf S},{\bf S}') \equiv \sum_i (1-\delta_{S_i,S'_i})
\ee
where $\delta$ is the Kronecker delta ($\delta_{S_i,S'_i}=1$ if $S_i=S'_i$, $\delta_{S_i,S'_i}=0$ if $S_i\neq S'_i$). 
Given a reference state ${\bf S}_\text{ref}$, the Hamming distance distribution (over time) is given by
\be
\prob(HD,t)=\sum_{\bf S}\prob\left[{s}(t)={\bf S}\right]\delta_{HD,HD({\bf S},{\bf S}_\text{ref})}
\ee

The estimation of discrete time Hamming distance distribution $\prob(HD,\tau)$ is similar to that of stochastic probabilities on time window.

The Hamming distance distribution is a useful characterization when the set of instantaneous probabilities is compared to a reference state (${\bf S}_\text{ref}$). In that case, Hamming distance distribution describes how far this set is to this reference state. 

The Hamming distance distribution can be considered as a ``semi-global'' characterization of time evolution: for a given time window, the size of this characterization is the number of nodes (to be compared with probabilities on time window, whose size is $2^\text{\#nodes}$).


\subsubsection*{\label{sect:icoutputs} Initial conditions and outputs}
{\sl Inputs Nodes} are defined as the nodes for which the initial condition is fixed. Therefore, each trajectory of BKMC starts with fixed values of input nodes and random values of other nodes.

{\sl Internal nodes} are nodes that are not considered for computing probability distributions, entropies and transition entropies. {\sl Output nodes} are nodes that are not internal. Technically, probabilities are sum up over network states that differ only by the state of internal nodes. These internal nodes are only used for generating time trajectories through BKMC algorithm.
Usually, nodes are chosen to be internal when the corresponding species is not measured experimentally. 
Mathematically, it is equivalent to transform the original Markov process to a new stochastic process (that is not necessary Markovian) defined to a new network state space; this new state space is defined by states of output nodes.
This raises the question of the transition entropy $TH$: formally, this notion has only a sense within Markovian processes, i.e. when there are no internal nodes. Here, we generalize the notion of transition entropy even in the case of internal nodes. Suppose that the system is in state ${\bf S}$:

\begin{itemize}
\item If the only possible transitions from state ${\bf S}$ to any other state consist of flipping an internal node, the transition entropy is zero.
\item If there is, at least, one transition from state ${\bf S}$ to another state that flips a non-internal node, then only the non-internal nodes will be considered for computing probabilities in equations \ref{eq:ProbTrans}. In particular, $\sum_{{\bf S}'}\rho_{{\bf S}\rightarrow{\bf S}'}$ is computed only on non-internal node flipping events.
\end{itemize}

{\sl Reference nodes} are nodes for which a reference node state is specified and for which the Hamming distance is computed. In this framework, a reference state is composed of reference nodes for which the node state is known and non-reference nodes for which the node state is unknown. Note that non-reference nodes may differ from internal nodes.   



\subsubsection*{Stationary distribution characterization}

It can be shown (see Supplementary Material, corollary 2, section 1.2) that instantaneous probabilities of a continuous time Markov process converge to a stationary distribution. Fixed points and cycles are two special cases of stationary distributions. They can be identified by the asymptotic behavior of entropy and transition entropy (this works only if no nodes are internal): 

\begin{itemize}
\item if the transition entropy and the entropy converge to zero, then the process converges to a fixed point 
\item if the transition entropy and the entropy does not converge to zero, then the process converges to a cycle 
\end{itemize}

More generally, the complete description of the Markov process asymptotic behavior can be expressed as a linear combination of the indecomposable stationary distributions. 

A set of finite trajectories, produced by BKMC, can be used to estimate the set of indecomposable stationary distributions.
Consider a trajectory $\hat{\bf S}(t), t\in [0,T], i=1,\cdots, n$. Let $I_{\bf S}(t) \equiv \delta_{{\bf S},\hat{\bf S}(t)}$. 
The estimation of the associated indecomposable stationary probability distribution  (${s}_0$) is done by averaging over the whole trajectory:
\be
\hat{\prob}\left[{s}_0={\bf S}\right]=\frac{1}{T}\int_0^TdtI_{\bf S}(t)
\ee

Therefore, a set of indecomposable stationary distribution estimates can be obtained by a set of trajectories. These indecomposable stationary distribution estimates should be clustered in groups, where each group consists of estimates for the same indecomposable stationary distribution. For that, we use the fact that two indecomposable stationary distributions are identical if they have the same support, \emph{i.e.} the same set of network states with non-zero probabilities (shown in Supplementary Material, theorem 2). Therefore, it is possible to quantify how similar two indecomposable stationary distribution estimates are. A {\sl similarity coefficient} $D({s}^{(i)}_0,{s}^{(j)}_0) \in [0,1]$, given two stationary distribution estimates ${s}^{(i)}_0$ and ${s}^{(j)}_0$ , is defined:  

\be
D({s}^{(i)}_0,{s}^{(j)}_0)\equiv
\left(\sum_{{\bf S}\in \text{support}({s}^{(i)}_0,{s}^{(j)}_0)}\hat{\prob}\left[{s}^{(i)}_0={\bf S}\right]
\right)
\left(\sum_{{\bf S}'\in \text{support}({s}^{(i)}_0,{s}^{(j)}_0)}\hat{\prob}\left[{s}^{(j)}_0={\bf S}'\right]
\right)
\ee
where
\be
\text{support}({s}^{(i)}_0,{s}^{(j)}_0)\equiv
\left\{{\bf S}\text{ such that }\hat{\prob}\left[{s}^{(i)}_0={\bf S}\right]\hat{\prob}\left[{s}^{(j)}_0={\bf S}\right]>0\right\}
\ee
 

Clusters can be constructed when a similarity threshold $\alpha$ is provided.
 A cluster of stationary distributions is defined as follows: 
\be
{\cal C}=\left\{{s}_0|\;\exists {s}'_0\in {\cal C} \text{ s. t. } 
D({s}_0,{s}'_0)\geq\alpha\right\}
\ee  

For each cluster ${\cal C}$, an associated distribution ${s}_{\cal C}$, that estimates an indecomposable stationary distribution, can be defined 
\be
{\prob}\left[{s}_{\cal C}={\bf S}\right]=\frac{1}{|{\cal C}|}\sum_{{s}\in {\cal C}}
{\prob}\left[{s}={\bf S}\right]
\ee

Errors on this estimate can be computed by 
\be
\text{Err}\left({\prob}\left[{s}_{\cal C}={\bf S}\right]\right)=\sqrt{\text{Var}({\prob}\left[{s}={\bf S}\right],{s}\in {\cal C})/|{\cal C}|}
\ee

Notice that this clustering procedure has no sense if the process is not Markovian; therefore, no node will be considered as internal. 




\section*{List of abbreviations}

BKMC: Boolean Kinetic Monte-Carlo \\
AT: Asynchronous transition \\
ODEs: Ordinary Differential Equations \\
MaBoSS: Markov Boolean Stochastic Simulator \\

\section*{Author's contributions}
G. Stoll organized the project, set up the algorithms, participated in writing the software, set up the examples and wrote the article.

E. Viara wrote the software and participated in setting up the algorithms.

E. Barillot participated in writing the article.

L. Calzone organized the project, set up the examples and wrote the article.

\section*{Acknowledgements}
  \ifthenelse{\boolean{publ}}{\small}{}
This project was supported by the Institut National du Cancer (SybEwing project), the Agence National de la Recherche (Calamar project).  The research leading to these results has received funding from the European Community's Seventh Framework Programme (FP7/2007-2013) under grant agreement nb HEALTH-F4-2007-200767 for APO-SYS and nb FP7-HEALTH-2010-259348  for ASSET. GS, EB and LC are members of the team ”Computational Systems
Biology of Cancer”, Equipe labellis\'ee par la Ligue Nationale Contre le
Cancer. We'd like to thank Camille Sabbah, Jacques Rougemont, Denis Thieffry, Elisabeth Remy, Luca Grieco and Andrei Zinovyev.
 

\newpage
{\ifthenelse{\boolean{publ}}{\footnotesize}{\small}
  \bibliography{bmc_article} }     


\ifthenelse{\boolean{publ}}{\end{multicols}}{}

\end{document}